\documentclass[runningheads,a4paper]{llncs}
\usepackage{geometry}
\usepackage{url}
\usepackage{threeparttable}
\usepackage{listings}
\usepackage{multirow}
\usepackage{amsmath}
\usepackage{amssymb}
\usepackage{fancybox, graphicx}
\usepackage{url}
\usepackage[table]{xcolor}
\usepackage{textcomp}
\usepackage{lscape}
\usepackage{booktabs} 
\usepackage{colortbl}
\usepackage{paralist}

\usepackage{hyperref}
\hypersetup{
  colorlinks=true,
  linkcolor=black,
  urlcolor=blue!70!black,
  citecolor=blue!70!black,
}

\usepackage{cite}
\usepackage{eurosym}
\usepackage{graphicx, subfigure}

\usepackage{wrapfig}

\usepackage{color}

\begin{document}

\mainmatter  

\title{Cyber-Physical Architecture Assisted by Programmable Networking}

\titlerunning{Cyber-Physical Architecture Assisted by Programmable Networking}

\authorrunning{Rubio-Hernan, Sahay, De Cicco, Garcia-Alfaro}

%
%
\author{Jose Rubio-Hernan$^1$ \and Rishikesh Sahay$^2$ \and Luca De Cicco$^3$ \and Joaquin Garcia-Alfaro$^1$}

%

\institute{$^1$~Institut Mines-T\'{e}l\'{e}com, T\'{e}l\'{e}com SudParis, France\\
$^2$~Technical University of Denmark, Denmark\\
$^3$~Politecnico di Bari, Italy
}
%
%

\maketitle

\begin{abstract}

Cyber-physical technologies are prone to attacks, in addition to faults and failures. The issue of protecting cyber-physical systems should be tackled by jointly addressing security at both cyber  and physical domains, in order to promptly detect and mitigate cyber-physical threats. Towards this end, this letter proposes a new architecture combining control-theoretic solutions together with programmable networking techniques to jointly handle crucial threats to cyber-physical systems. The architecture paves the way for new interesting techniques, research directions, and challenges which we discuss in our work.\\


\end{abstract}

\section{Introduction}
\label{sec:intro}

Cyber-physical systems integrate physical infrastructures over computing network resources, in an effort to reduce complexity and costs of traditional control systems. Modern industrial control systems are a proper example. They have evolved from networked-control systems that route their underlying information in terms of physical measurement and feedback control. Traditional security, from a network and computing security standpoint, is able to cover cyber threats, but fails at addressing physical threats. Control-theoretic solutions, combined with network computing security techniques, can lead to powerful solutions to cover both physical and cyber-physical attacks at the same time~\cite{pasqualetti2015control,Mo_2015,watermark_ETT}. Nevertheless, guaranteeing the resilience of these systems (i.e., to keep offering critical functionality under attack conditions) is still an open and critical problem to solve~\cite{urbina2016survey}. We argue that the use of \textit{Programmable Networking} is a plausible solution to efficiently handle such a problem. 

Programmable networking is a paradigm to manage network resources in a programmatic way. It can be used to decouple static network architectures and ease their management. Additionally, it facilitates the decentralization of network control and processing of network data. Control and data domains can be reprogrammed dynamically to enforce novel network applications and functionalities. Software-Defined Networking (SDN)~\cite{Kreutz_SDN_survey} is a promising technology associated to the concept of programmable networking.  The use of programmable networking allows decoupling the control domain from the data domain~\cite{onf12}. Novel network functionality can be devised and deployed depending on the current requirements. This includes the enforcement of protection schemes to recover the system elements from attacks~\cite{fresco_Shin,bohatei,drawbridge}. The use of programmable networking increases the  visibility of controllers in terms of failure and attacks. It enables network operators to take control over the network flows based on dynamic conditions~\cite{hedera}. It also allows to control data domain devices from application-layer controllers, hence increasing the flexibility in terms of network reconfiguration. For instance, applications at the control domain can dynamically deploy policies over the networked-devices via high-level programming languages~\cite{aroma}.

In this letter, we present a novel approach towards the development of a programmable cyber-physical system. Network and physical controllers get connected towards coordinating resilience strategies, e.g., to maintain the resilient properties of the system under failure and attacks, at any layer of the resulting architecture. The proposed architecture   is experimentally validated by means of current programmable and cyber-physical technologies. More specifically, we show a proof-of-concept design combining SCADA (Supervisory Control And Data Acquisition) protocols and SDN-enhanced topologies, in order to evaluate the improvement of detection and mitigation capabilities of the resulting system.




The remainder sections of this letter are organized as follows. Section~\ref{sec:our-proposal} presents our envisioned architecture. Section \ref{sec:discussion} discusses about the advantages and limitations of our proposal, w.r.t. experimental feasibility of our work. Section~\ref{sec:conclusion} closes the letter with some conclusions and perspectives of future work.

\section{Our Proposal}
\label{sec:our-proposal}

\subsection{Combining two complementary paradigms}

\begin{figure}[!b]
\centering
   \subfigure[Proposed architecture, combining feedback control and programmable networking]{
   \label{fig:cpspn}
     \includegraphics[width=10cm]{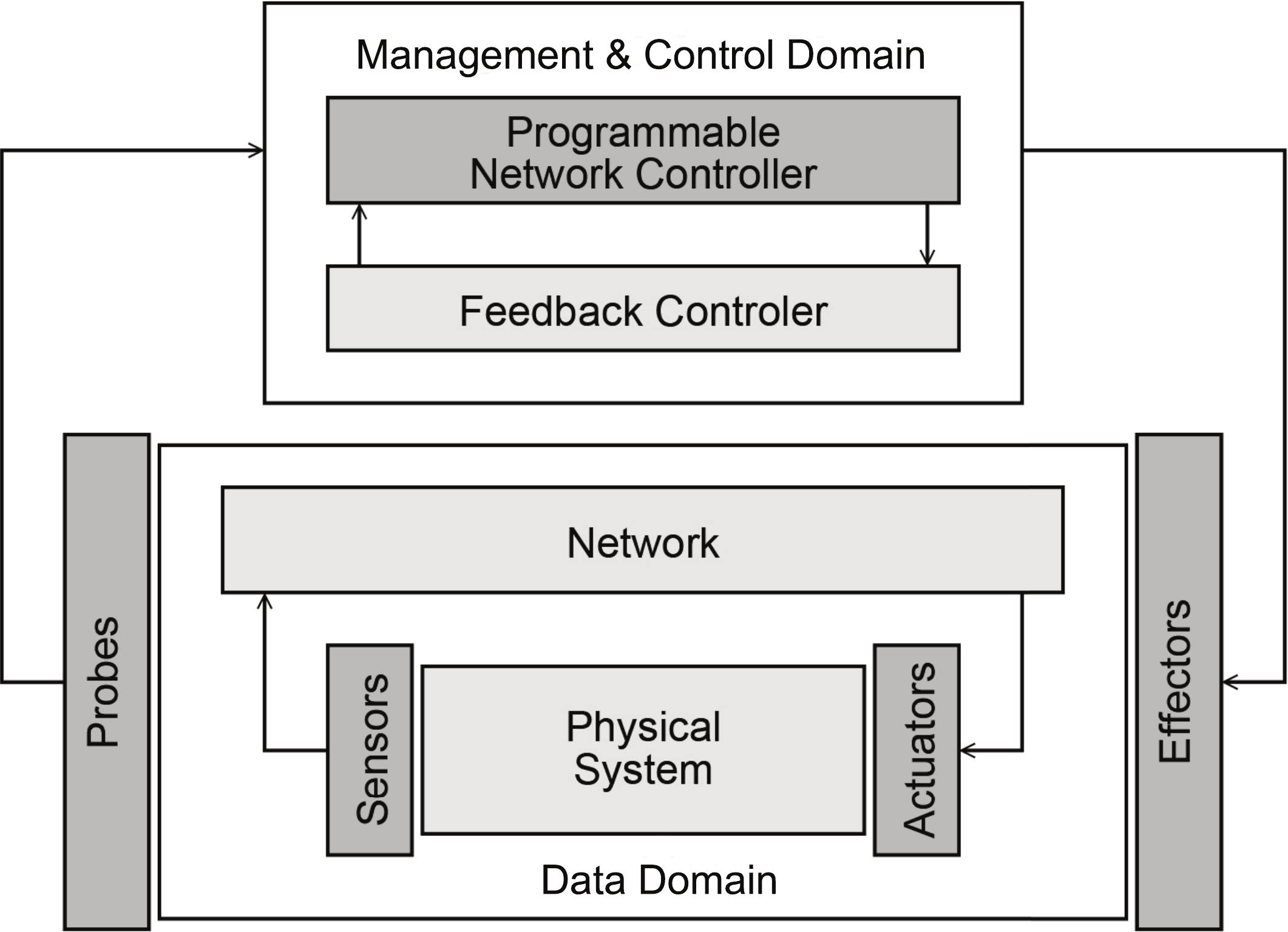}
   }
   \subfigure[Closed-loop feedback control]{
   \label{fig:cps}
     \includegraphics[width=5cm]{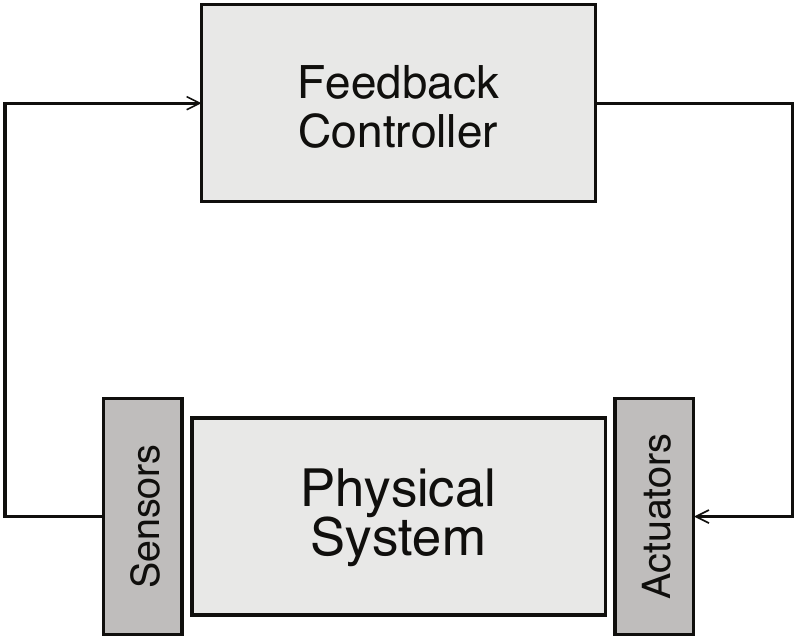}
   }
   \subfigure[Programmable networking]{
   \label{fig:pn}
     \includegraphics[width=5cm]{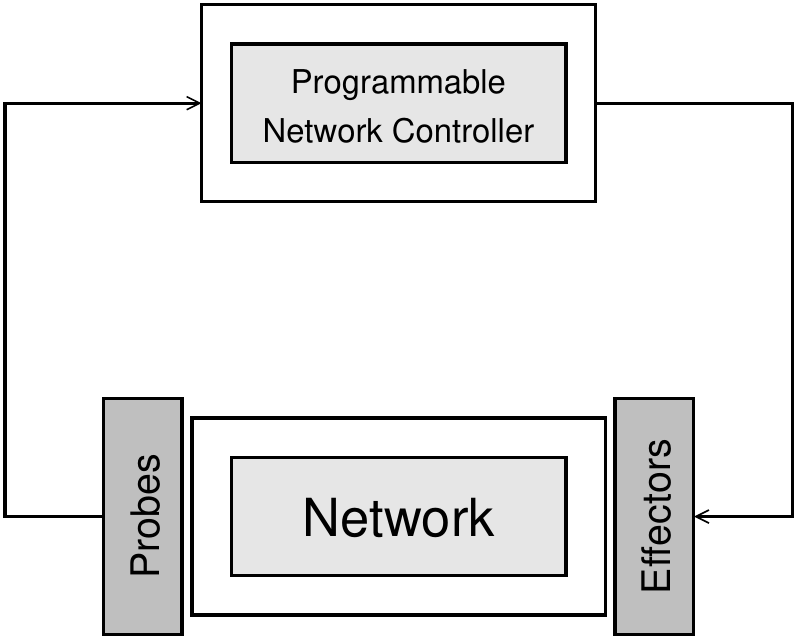}
   }
\caption{Proposed architecture}
\label{fig:figures1-2-3}
\end{figure}

\textit{Cyber-physical systems} and \textit{programmable networking} are two complementary paradigms, but often separately addressed by two different research communities (control and computing-network communities). Both paradigms use similar elements, that can be presented either following control-theoretic architectures~\cite{Lindberg_CPS} or via  computer-based programmatic definitions~\cite{matni2015technical}. In the control community, cyber-physical systems are regarded as a particular class of networked-control systems~\cite{Lindberg_CPS} (cf. Figure~\ref{fig:figures1-2-3}). Particularly, feedback control is managed by the following elements. {\it Controllers}, located within the cyber layer of the system (i.e., the layer related to the network and computing resources), monitor and supervise information produced by physical {\it sensors} reporting measurements from physical processes. Based on the sensor measurements, controllers dynamically compute corrective actions which are put in place by system {\it actuators}, to steer the physical processes to the desired states. 

Programmable networking can be represented using similar elements and definitions~\cite{matni2015technical}, as depicted in Figure~\ref{fig:figures1-2-3}. In such a representation, the controller is governed by software elements, supervising both the {\it management} and the {\it control} domains of a programmable architecture. The controller manages the behavior of all the interactions at the {\it data domain}, where network elements (e.g., network switches) follow some precise policies directed by the controller (e.g., to enforce routing and security strategies). The remainder elements of the network get permanently monitored by the controller, and orchestrated via control policies to improve, e.g., the routing of traffic or the enforcement of network security countermeasures. To conduct monitoring at the data domain, several network probes (referred to as meters in Figure~\ref{fig:figures1-2-3}) report networking measurements to the controller.

\subsection{Towards resilient control systems}

A crucial goal of both cyber-physical designs and programmable networking is to ensure resilient control systems. Resilience is a property that has been studied for many years in different areas of knowledge. Laprie \cite{laprie208} defines the resilience of a system as {\it the persistence of dependability when facing changes in the system}.
Dependability has also been expressed by Ellison \textit~{et al.} as the {\it avoidance of failures that are unacceptably frequent or severe}~\cite{ellison1997survivable}. From these two definitions, we see that resilience deals with the management of operational functionality that is crucial for a system, and that cannot be stopped. In other words, system functionality that shall be properly accomplished. For instance, the cooling service of reactor in a nuclear plant is a proper example of a critical functionality.

Other system functionalities may be seen as less important. They can
even be temporarily stopped or partially accomplished. Such type of
functions can be seen as secondary functions. A printing service for
employees in the aforementioned nuclear plant scenario is a proper
example of a secondary function. Another important element to take into consideration is the severity of failures. The more severe a failure is, the more it will affect the related system. Also, the more severe the failure, the harder will be for the system to recover the nominal system functionalities. Several works have focused on these issues when addressing the survivability of critical system functionalities (see, for instance, reference \cite{linger1998requirements} and citations thereof).

Finally, a resilient control system shall able to
\cite{queiroz2012holistic}: (i)~detect undesirable threats; (ii)~minimize the impact of such threats; and (iii)~mitigate them (i.e., recover to normal operation in a reasonable time). Solutions include the use of redundancy (e.g., use of software replicas), the enforcement of  system segmentation, and the inclusion of different levels of protection (e.g., secure handshakes, message signatures, and encryption). In the end, the goal is to include enough security measures to closing the system monitoring loop with additional responses that mitigate the undesirable events. In other words, inclusion of mitigation techniques to keep processes in a normal operating state while the system is confronted to failures and attacks. The challenge of satisfying those previous requirements on a cyber-physical system is the difficulty of properly differentiating failures from attacks, since their underlying security models (e.g., in terms of mitigation) differ as well.

\subsection{Cyber-Physical Architecture Assisted by Programmable Networking}
\label{sec:proposal}

Taking into account the aforementioned descriptions and goals, in this section we propose a resilient cyber-physical architecture assisted by programmable networking. The proposed architecture allows creating a cross control layer between the physical and the cyber layers. The resulting design aims to enable the combination of different security models, e.g., from a security and safety standpoint. For instance, to enable adaptive mitigation of threats, making possible to differentiate between faults, failures and attacks, prior enforcing the eventual mitigation strategies. The use of programmable networking allows to move to a higher level of abstraction when analyzing the system threats (in contrast to traditional solutions anchored at the data domain), moving the defense at the same level of cyber-physical adversaries, assumed to be entities with equivalent powers (e.g., in terms of observability and controllability).


Figure~\ref{fig:cpspn} shows our proposed framework of a programmable networking assisted cyber-physical system. The framework contains different components of programmable networking (PN) technologies and cyber-physical systems (CPS): (1)~The data domain is mainly comprised of two sub-spaces: A physical space composed by {\it physical sensors and actuators}, which are used by the CPS to communicate with the physical processes located in the physical system (also located in the physical space). These devices communicate with the Feedback controller to manage the physical processes; And a network space composed by {\it PN switches}, which are programmable network switches controlled dynamically. This dynamic control enables us to minimize the deployment cost of the network framework, as to improve and manage many other network features~\cite{Sharma_thesis}, e.g., QoS and security requirements. PN switches use a centralized framework with an interface to control and manage all the network features dynamically; (2)~The management and control domain contains two different controllers working on a joint and coordinated way, to fully cover the control of the resulting framework. The  controllers are: (i)~a {\it Feedback controller} and (ii)~a {\it Programmable Network (PN) controller}. 

The Feedback controller is made of two sub-components: (a) A feedback controller that is in charge of enforcing the dynamical control objectives (fast dynamics are involved); (b) A supervisor controller that communicates in a bi-directional way with the PN controller in the following way. The PN controller, based on measurements provided by probes and feedback provided by the Feedback controller, is able to detect a possible threat acting on the control path. In response to such a threat, the PN controller provides a corrective measure to mitigate the impact. The PN controller can be seen as a computing entity that is located at an external location (e.g., a kind of Network Operating System~\cite{Kreutz_SDN_survey}). For instance, it provides resources and abstractions to manage the system using a centralized or a decentralized model~\cite{Nordsec}. 


Together, both controllers manage the data domain. Feedback
controller manages the physical system through physical sensors
and actuators deployed at the physical layer. And the PN
controller estimates and manages the data domain through network
probes and effectors ---deployed at the management and control
domain. It is worth noting that the PN controller uses system
identification tools to estimate the behavior matrices of the
physical system. As a result, it can compare the {\it nominal
cyber-physical system model} (i.e., the behavior matrices) with
the model estimated by the Feedback controller. The correlation
between both estimations allows to detect anomalies in the
system~(cf. reference~\cite{Nordsec}, for a sample technique
based on the same principle). Notice that such anomalies can
either be unintentional failures, or stealthy attacks disguised
by intentional modifications produced by a malicious entity who
is willing to disrupt the physical system
\cite{watermark_Eurasip,watermark_ETT}. Assuming a system with
appropriate measure to detect both failures and attacks, the PN
controller can react to those anomalies, by enforcing some {\it
mitigation} policies. At the data domain, we have {\it network
probes and effectors}, conducting data monitoring ---if
instructed by the control domain. Network probes monitor the
traffic in the data domain and provide the information to the PN
controller. The PN controller analyses the information and
forwards control actions to the effectors. Network rules at the
control domain are responsible to enforcing such actions. For
instance, when a network probe finds tampered traffic in a
network path, it provides the tampered information to the control
domain. Then, the PN controller, located at the control domain,
checks for the available resources in, e.g., a path lookup
component, and provides new routes to enforce the action. For
instance, it may redirect the tampered traffic to provide fair
share of network bandwidth w.r.t. legitimate traffic.

Another important component of the management and control domain
is the {\it Path Lookup}. Paths in the framework are
pre-computed. Normally, a list of paths are maintained at the
control domain. The path lookup component maintains a table of
paths (from the ingress switch to the egress switch) sorted
according to the quality of service provided by the paths,
associated to unique labels. Paths are later assigned to flows
based on the traffic class that they belong to. This is a crucial
component of the overall system. Resiliency to attacks can be
implemented by using multi-path communication between
cyber-physical system components or by activating new paths (even
suboptimal) to evade an attack. For example, legitimate flows are
assigned to high priority paths while suspicious flows are
assigned to paths containing middleboxes or paths having low
bandwidth and longer in terms of hops, to finally forward the
malicious flows through paths which lead to a sinkhole.



\section{Validation and Discussion}
\label{sec:discussion}

To validate the feasibility of our proposal, we are currently implementing a proof-of-concept prototype based on the architecture proposed in this letter. The prototype combines the components of a cyber-physical system together with programmable networking technology. It builds upon {\it Mininet}~\cite{mininet}, a network emulator tool which provides a rapid prototyping for programmable networking. The protocol used to instantiate the programmable networking techniques is OpenFlow. It allows controlling the path of the network traffic through programmable networking switches. For the implementation of cyber-physical system, we use the SCADA (Supervisory Control And Data Acquisition) Modbus protocol~\cite{modbusspecs}. Further information about the proof-of-concept prototype and ongoing results is \href{http://j.mp/TSPCPSDN}{available online}. 

For the time being, the programmable networking controller of the prototype  instructs the network probes of a cyber-physical system (e.g., autonomous movable vehicles) to monitor traffic from data domains, in order to investigate the existence of malicious traffic evidences. More specifically, the programmable networking controller receives traffic header details (such as source IP address, destination IP address and protocol) along with payload data containing networked-feedback measurements from the physical domains. The programmable networking controller cooperates with the Feedback controller. It analyzes anomalous details and  may shares information with the Feedback controller. Whenever the level of suspicious events reaches a configurable threshold, the programmable networking controller may decide to send the measurements to, e.g., quarantine subnetwork, by redirecting the suspicious traffic through alternative routing paths \cite{hadega}. This solution, based on the path lookup module of {\it Mininet}, uses network traffic labeling to mark down the suspicious events. This is implemented in practice by adding an additional functionality to the {\it Mininet} programmable networking controllers that deploys mitigation rules to handle suspicious traffic at the edge of the programmable networking switches of the cyber-physical system. The redirection of traffic is enforced by the network effector of the system, in order to enable quarantining plans. Some more practical details and video captures of the proof-of-concept prototype are available online at \url{http://j.mp/TSPCPSDN}.

The aforementioned prototype validates the architecture proposed in this letter. It promotes cooperation between controllers located at both management and control domains of a cyber-physical system. Nevertheless, further analysis is required w.r.t. the duality of goals of each controller paradigm represented in the architecture depicted in Figure~\ref{fig:cpspn}. In some cases, the two combined paradigms may be driven with different objectives. A more thorough study shall be conducted to identify how the controllers achieve a common goal (e.g., guaranteeing  cyber-resilience of the underlying system) withing interfering their primary objectives (i.e., networking and physical control objectives). To make a concrete example, the Feedback controller can report  an anomalous deviation of the sensor readings from the nominal behavior as a feedback to the PN controller. This issue could be due to sensors being faulty or due to an attack whose surface intersects with the control paths. Unfortunately, the Feedback controller alone is not able to distinguish between the two events since it has no view of the underlying network (information hiding). In response to such a situation, the Feedback controller sends an alert signal to the programmable networking controller which verifies if the control path (which is unknown to the Feedback controller) is likely to be under attack. In case of an attack is detected by the programmable networking controller, it puts in place the corrective measures (i.e., segregate malicious traffic programmatically) and sends a signal to the Feedback controller, to report that a corrective action has been taken into account. Notice that sophisticated attacks to the system could escape the detectors running solely in the cyber-layer. When this is managed as well by the Feedback controller, stealthy attacks hidden as failure and faults will be identified. A more precise evaluation of this type of scenarios will be provided in a forthcoming publication.

\section{Conclusion}\label{sec:conclusion}

This letter shows that programmable networking and feedback control can be combined together in order to build higher resilient cyber-physical systems. We argued that the construction a programmable networking-assisted  cyber-physical architecture, improves detection and mitigation of cyber-physical attacks. It allows cooperation between traditional Feedback controllers and programmable networking devices, to allow more efficient mitigation of threats (e.g., by providing evidences about stealthy cyber-physical attacks disguised as failures and faults). Next steps include a more thorough analysis about the cooperation of controllers, as well as investigation about the effective activation of cyber-physical resilience by combining novel detection and mitigation strategies. 









\section*{Acknowledgment}
The research in this paper has received partial funding from the Cyber-CNI Chair of the Institut Mines-Telecom (\url{http://chaire-cyber-cni.fr/}).

\bibliographystyle{abbrv}
\bibliography{main}  

\end{document}